\documentclass{article}
\usepackage{spconf,amsmath,graphicx}
\usepackage{algorithm,algorithmic}
\usepackage{cite}
\usepackage{amssymb,amsfonts}
\usepackage{graphicx}
\usepackage{textcomp}
\usepackage{multicol}
\usepackage{xcolor}
\usepackage[mathscr]{euscript}
\usepackage{xcolor}
\usepackage{soul}
\usepackage{csquotes}
\usepackage{svg}
\usepackage{adjustbox}
\usepackage{url}
\newcommand{\Imgwidth}{99pt}
\title{Balanced Deep CCA for Bird Vocalization Detection}
\ninept
\name{Sumit Kumar\textsuperscript{1}, B. Anshuman\textsuperscript{1},
Linus R{\"u}ttimann\textsuperscript{2},
Richard H.R. Hahnloser\textsuperscript{2}, Vipul Arora\textsuperscript{1}}
\address{\textsuperscript{1}Department of Electrical Engineering, Indian Institute of Technology, Kanpur, India.\\
\textsuperscript{2}Institute of Neuroinformatics, University of Zürich and ETH Zürich, 8057 Zürich, Switzerland.}

\begin{document}
\maketitle
\begin{abstract}
Event detection improves when events are captured by two different modalities rather than just one. But to train detection systems on multiple modalities is challenging, in particular when there is abundance of unlabelled data but limited amounts of labeled data. We develop a novel self-supervised learning technique for multi-modal data that learns (hidden) correlations between simultaneously recorded microphone (sound) signals and accelerometer (body vibration) signals. The key objective of this work is to learn useful embeddings associated with high performance in downstream event detection tasks when labeled data is scarce and the audio events of interest --- songbird vocalizations --- are sparse. We base our approach on deep canonical correlation analysis (DCCA) that suffers from event sparseness. We overcome the sparseness of positive labels by first learning a data sampling model from the labelled data and by applying DCCA on the output it produces. This method that we term balanced DCCA (b-DCCA) improves the performance of the unsupervised embeddings on the down-stream supervised audio detection task compared to classsical DCCA. Because  data labels are frequently imbalanced, our method might be of broad utility in low-resource scenarios.

\begin{keywords}
self-supervised learning, representation learning, multi-modal, DCCA, data imbalance
\end{keywords}

\end{abstract}
\section{Introduction}
\label{sec:intro}

Supervised learning techniques have produced very promising results in speech signal processing. But supervised learning tends to be inadequate in low-resource fields such as animal communication research where large datasets of labelled vocalizations are very time-consuming and costly to produce. In such Sound Event detection (SED) tasks, both the event classes and the events' timelines (onsets and offsets) need to be labeled, which is a tedious task in noisy environments. In contrast, in self-supervised learning (SSL), models themselves generate supervisory labels from unlabeled data, with the benefit of reducing the dependence on supervisory labels. By learning from unlabeled data, unsupervised models leverage the underlying data structure.

Self-supervised learning on multi-modal data can be implemented on Siamese Networks \cite{b1}. These networks are generally trained using a contrastive loss \cite{b2}, \cite{b3}. In their objective function, Jure et al. \cite{b4} used the correlation between two distorted views of the same image as distance metric.
In natural language processing, word2vec \cite{b5}, BERT \cite{b6} are popular techniques for learning the relationship between words in a self-supervised manner; these techniques are based on the concept of masking.
Recently, the field of audio signal processing has adopted SSL because of its success in computer vision and natural language processing. Inspired by word2vec, Steffen  et al. proposed wav2vec \cite{b7}. In the case of multi-view or multi-modal data, Wang et al. used Deep Canonical Correlation Analysis (DCCA) \cite{b8} to learn the correlated acoustic features \cite{b9}. In \cite{iccn}, the authors extracted the (hidden) correlation between the audio, video, and text features using the proposed ICCN network. In \cite{b_fuse}, the correlated features obtained from DCCA are used for the emotion recognition task. The obtained features are fused in different ways and used for various downstream tasks \cite{b10}
We aim to apply SSL for SED task using multi-modal data. DCCA can be used to leverage the underlying correlation between different modalities of data. The dataset used for this task is highly sparse. Therefore, DCCA-based approaches fail to perform. To address this issue, we propose a novel balanced deep canonical correlation analysis (b-DCCA) model where the DCCA model is balanced over the classes. Our model requires a very limited amount of labeled data in order to bootstrap the DCCA model. The labeled data used for the downstream task can be used for bootstrapping the DCCA model. The main contributions of this paper are summarized as follows.
\begin{enumerate}
    \item This paper proposes a novel self-supervised learning model called balanced deep canonical correlation analysis (b-DCCA) that deals with sparse datasets by maximizing the entropy of training batches through a binning technique.
    \item We also show the proposed b-DCCA model trained on both modalities of the data may require only a single modality at the time of inference, thus further reducing the cost of data collection.
    \item We release a dataset named \textbf{TwoRadioBirds} for birds vocalization detection task. The dataset can be accessed using the link: \url{https://doi.org/10.5281/zenodo.7253729}
\end{enumerate}
Our codes are available on \url{https://github.com/madhavlab/2022_icassp_krsumit} 

\section{Dataset}
\label{sec:dataset1}
The bird vocalization dataset \textbf{TwoRadioBirds} contains accelerometer and microphone recordings of an adult male and an adult female zebra finch (\emph{Taeniopygia guttata}) that were housed in a homecage were they could freely behave. All experimental procedures were approved by the Cantonal Veterinary Office of the Canton of Zurich, Switzerland (license numbers ZH045/2017). All methods were carried out in accordance with relevant guidelines and regulations (Swiss Animal Welfare Act and Ordinance, TSchG, TSchV, TVV). Vibration transmitter devices were mounted on the back of the birds with a rubber band harness \cite{ruttimann2022multimodal}. The vibration transducer (Knowles Bu-21771-000 accelerometer) on the transmitter functioned as a contact microphone and selectively recorded vocalizations of the bird that carried the transmitter. The sensor signal was routed into a high-pass filter (-3dB: 15Hz) and then transmitted as a frequency modulated FM radio signal. The vibration transducer signal picks up frequencies from vocalizations up to 7kHz in the best case and up to 1kHz in the worst case, depending on factors like e.g. how well the skin contact is. Additionally the transmitter devices picks up movement signals from e.g. wing flaps and radio noises \cite{ruttimann2022multimodal}. An additional, modified transmitter device, where the vibration transducer was replaced by a Knowles FG-23329-D65 microphone, was mounted on the wall of the homecage, it delivered the microphone signal. The radio signals from the vibration and microphone transmitter device were received and demodulated by a software defined radio receiver.

The dataset was annotated by visual inspection of accelerometer vibration spectrograms, which was separately performed for the male and the female vibration data. In the first step, candidate vocalizations were extracted as described in \cite{rich}. In the second step, the candidate vocalizations were manually corrected by discarding them as noise or by manually shifting the vocalization onset and offset times to their correct positions. In a third step, we looked for missed vocalizations using the positive examples thus far obtained as templates and by performing a brute-force nearest neighbor search across the entire dataset of candidate vocal segments (the latter were defined as the regions of supra-threshold vibration signal amplitudes, excluding the already-extracted vocalizations). The dataset consists of eleven data files of 1 hr. duration each, out of which three are densely annotated and eight are unlabelled. Each file contains data from a microphone (audio) and two accelerometers (body vibration) channels. 
The data in all channels is sampled at a rate of 24000 Hz.
\begin{figure}[h]
    \centering
    \includegraphics[width=0.3\textwidth]{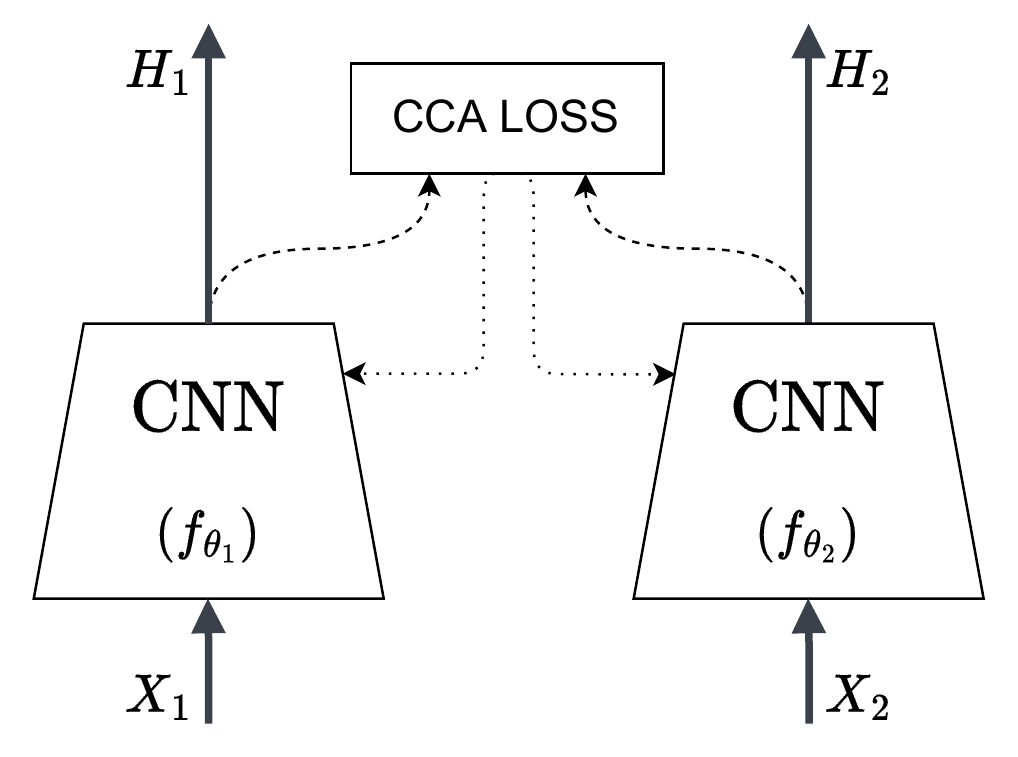}
    \caption{DCCA model}
    \label{dcca_fig}
\end{figure}

\section{Deep Canonical Correlation Analysis (DCCA)}
\label{Deep Canonical Correlation Analysis}
CCA~\cite{b0} is a standard statistical technique for finding linear projections of two random vectors such that the projections are maximally correlated.
Given two random vectors $(X_1,X_2) \in \mathbb{R}^{n_1}\times\mathbb{R}^{n_2} $ with covariances $(C_{11},C_{22})$ and cross-covariance $C_{12}$, CCA finds pairs of linear projections of the two views, $(w'_1X_1,w'_2X_2)$ that are maximally correlated:
\begin{equation}
 \begin{aligned}
\label{cca}
  (w_1^*,w^*_2) &=  \underset{(w_1 \in \mathbb{R}^{n_1}, w_2 \in \mathbb{R}^{n_2})}{\text{argmax}}~~ {\text{corr}(w'_1X_1,w'_2X_2)} \\
  &= \underset{(w_1 \in \mathbb{R}^{n_1}, w_2 \in \mathbb{R}^{n_2})}{\text{argmax}}~~ \frac{{w'_1\Sigma_{12}w_2}}{\sqrt{w'_1\Sigma_{11}w_1w'_2\Sigma_{22}w_2}}
  \end{aligned}   
\end{equation}

Both the projections are constrained to have unit variance. So eq.~\ref{cca} reduces to
\begin{equation}
\label{cca_obj}
    (w_1^*,w^*_2) = \underset{w'_1\Sigma_{11}w_1=w'_2\Sigma_{22}w_2=1}{\text{argmax}}{w'_1\Sigma_{12}w_2}
\end{equation}

The obtained pair $(w_1^{(1)},w^{(1)}_2)$ is known as first pair of canonical variables. Subsequent pairs are obtained sequentially by constraining each pair of projections to be uncorrelated with the previous one, i.e., $w^i_1\Sigma_{11}w^j_1=w^i_2\Sigma_{22}w^j_2=0$ for $i<j$. Stacking top $k ~ \text{projection vectors}~w^i_1$ as columns of matrix $A_1$ such that  $A_1 \in \mathbb{R}^{(n_1 \times k)},
$ ~similarly~ $ A_2 \in \mathbb{R}^{(n_2 \times k)}$.
As given in \cite{book}, the optimal solution of the CCA objective as defined in eq.~\ref{cca_obj} is given as $(A^*_1,A^*_2) = (C_{11}^{-1/2}U_k,C_{22}^{-1/2}V_k)$, where $U_k$ and $V_k$ are left and right singular matrices of $T, ~\text{and}~ T = \hat{C}_{11}^{-1/2}\hat{C}_{12}\hat{C}_{22}^{-1/2}$. The linear projections are given as $H_1 ~\text{and}~ H_2 \in \mathbb{R}^k$, where $H_1 = X_1^{'}A_1 ~\text{and}~ H_2 = X_2^{'}A_2$.

CCA can compute only linear projections which certainly limits its utility. Considering this limitation, Andrew et al. developed DCCA \cite{b9} which utilizes neural networks to learn the non-linear transformations of two views of data such that the embeddings obtained are highly correlated.
Data from both of the views is transformed by passing through several nonlinear layers. Let $(\theta_1, \theta_2)$ be the parameters of networks $(f_{\theta_1}, f_{\theta_2})$ as shown in fig.~\ref{dcca_fig}. For $m$ number of samples, $H_1 = f_{\theta_1}(X_1), H_2 = f_{\theta_2}(X_2) \in \mathbb{R}^{(o \times m)}$ are the outputs of DCCA network where $o$ is the number of neurons in the final layer of the network. Both $H_1 ~\text{and}~ H_2$ are centered about mean.  $\hat{C}_{11} = \frac{1}{m-1}H_1 H'_1 + r_{1}I, ~\hat{C}_{12} = \frac{1}{m-1}H_1 H'_2, ~\hat{C}_{22} = \frac{1}{m-1}H_2 H'_2 + r_{1}I$ are the estimated covariance matrices for both the views. A regularizer $r_1 > 0$ is added to ensure that the estimated matrices are positive definite. Summing up top $k$ singular values of $T$ matrix gives the total correlation of top $k$ components of $H_1 ~\text{and}~ H_2$. If $k=o$, the objective of DCCA is
\begin{equation}
\label{ccaloss}
    \text{corr}(H_1, H_2) = \lVert T \rVert_{tr} = \text{trace}(T'T)^\frac{1}{2}
\end{equation}
where $\lVert T \rVert_{tr}$ represents trace norm of $T$. The parameters $(\theta_1, \theta_2)$ are iteratively updated using the objective defined in eq. \ref{ccaloss}





\section{Data Preprocessing}
We merge the data from the male and female bird's accelerometer channels to create the first view of the DCCA, with the data from the microphone channel serving as the second view. For merging the data, we take the average over both the accelerometer channels. Now, having two channels, i.e., microphone and accelerometer, the labeled dataset is represented as $\mathcal{D}_l = \{(X^m,X^a,y)\}$, ~\text{and unlabeled as}~ $\mathcal{D}_u = \{(X^m,X^a)\}$. Superscripts $m$ and $a$ are used to denote data from microphone and accelerometer channels respectively throughout the paper.

Each file  in the dataset is segmented into clips of smaller duration. For each clip, the spectrogram $\tilde{X}\in\mathbb{R}^{f \times t}$ is calculated by squaring the magnitude of the spectrogram obtained by applying the short-time Fourier transform. $f$ and $t$ are the number of frequency bins and the number of time frames, respectively. Each time-frame $t$ of the spectrogram $\Tilde{X} \in \mathcal{D}_l$ is labeled using a binary value $b$, where $b \in \{0,1\}$. The frames with the label 1 indicate the presence of a sound event in that particular time frame and those with the label 0 indicate the absence of the sound event.
Since the dataset is highly sparse, class imbalance makes the task more challenging. For labeled dataset, the class imbalance is minimized by data augmentation technique, SpecAugment \cite{Specaugment}.

\section{Proposed Methodology}

\subsection{Module 1: Supervised Learning Module}
\label{SLM}
\begin{figure}[ht]
    \centering 
    \includegraphics[width=90pt, angle=90]{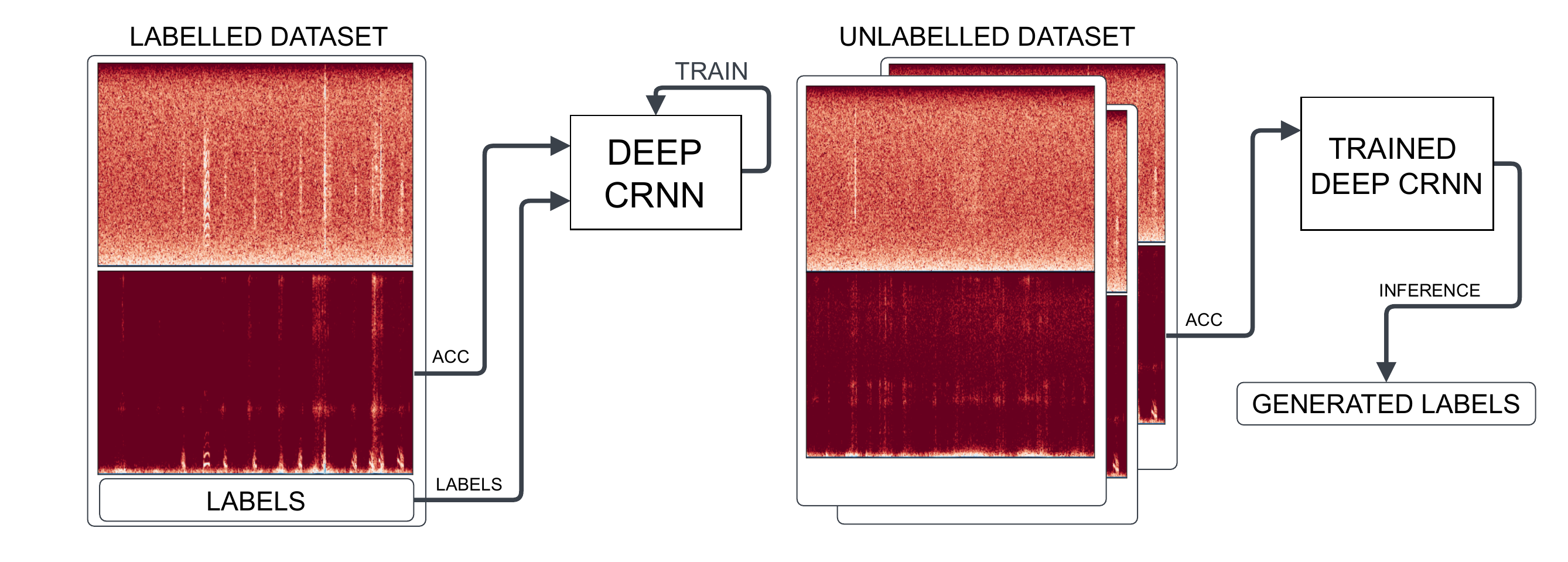}
    \caption{ Module 1: DCRNN model training and inference}
    \label{fig:module1}
\end{figure}

Deep convolutional recurrent neural network (DCRNN) \cite{cls} is used for the downstream task, i.e, birds vocalization detection.
Let $f_{\theta}$ be the binary classification DCRNN model where the model parameters $\theta$ are randomly initialized.
From $\mathcal{D}_l$, the spectrograms of the accelerometer channel, $\Tilde{X}^{a} \in \mathbb{R}^{f \times t}$ are given 
as an input to the model for bootstrapping the DCCA network as explained in following Section \ref{mod2}. For detection of birds vocalization, microphone channel of labeled dataset is used. For the given input, the model predicts $\hat{y} \in \mathbb{R}^t$.


The parameters $\theta$ are updated using the gradient descent algorithm: $\theta \leftarrow \theta - \alpha \boldsymbol{\triangledown}_{\theta} L(f_{\theta}),$where $\alpha \in \mathbb{R^+}$ is the learning rate and $L$ is the binary cross entropy loss. The trained model is then used to generate labels for unlabeled data as shown in fig. \ref{fig:module1} in order to bootstrap the DCCA module.

\subsection{Module 2: Balanced Deep Canonical Correlation Analysis (b-DCCA)}
\label{mod2}
\begin{figure}[h]
    \centering
    \includegraphics[width=\Imgwidth, angle=90]{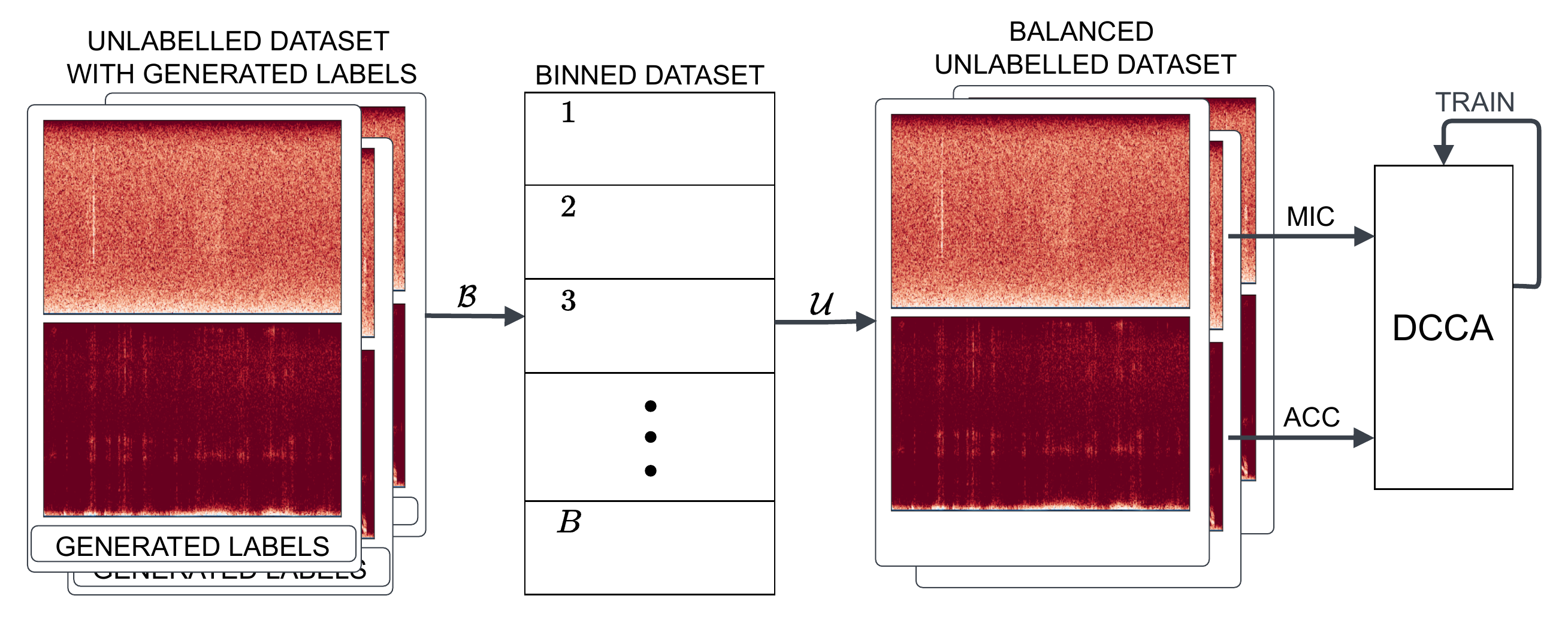}
    \caption{Module 2: Illustration of binning technique used in proposed b-DCCA model}
    \label{fig:module2}
\end{figure}
The model parameters of DCCA, $(\theta_1, \theta_2)$ as in discussed in Section \ref{Deep Canonical Correlation Analysis} are updated using the gradient descent algorithm using mini-batches. These mini-batches represent data distribution over all the classes present. For effective training, the mini-batches should contain data samples from each class with equal probability. Since the TwoRadioBird dataset is highly sparse, the majority class dominates over the minority class in most of the mini-batches used for training the model. Therefore, the model learns biased solutions. 
In order to solve this issue, the proposed b-DCCA algorithm, which maximises entropy across training batches over the classes of interest is discussed as follows.

First, the model $f_\theta$ from Module is trained on accelerometer data from labeled dataset. Then the trained $f_\theta$  is used for the inference on $\tilde{X}^a \in \mathcal{D}_u$. The predicted values obtained from $f_\theta$ are binarized using a unit-step function $U$ such that $\hat{y}^a = U(f_\theta(\tilde{X}^a)-\mathcal{T})$, where $\mathcal{T}$ is the threshold on the values of $f_\theta(\tilde{X}^a)$ and $\tilde{X}^a \in \mathcal{D}_u$. Value of $\mathcal{T}$ is chosen arbitrarily as 0.6. The values predicted by the model $f_\theta$ are an approximation of actual labels of accelerometer channel. Since, these values are just used for bootstrapping the DCCA model and not for the actual downstream task, i.e., birds vocalization detection, therefore the overall performance of the bird vocalization detection does not degrade.  Using the binarized $\hat{y}^ {a}$, the total number of sound events in a spectrogram can be calculated as $m = \sum_t \hat{y}^ {a}$ where $0 \leq m \leq t$. The value of $m$ is calculated for each spectrogram in $\mathcal{D}_u$. Out of all the calculated values of $m$, the maximum value is denoted by $M$. Further, the range $[0, M]$ is divided into $B$ equal parts. Based on the value of m, each spectrogram is allotted a bin index $n$ as defined below, 
\begin{equation}
\label{bin}
   n = \lceil \frac{m}{M} B \rceil
\end{equation}
where $n \in \{1, \cdots, B\}$.
The spectrograms are uniformly sampled from each bin to create training batches for b-DCCA as shown in fig.~\ref{fig:module2}. The spectrograms in the higher order bins have very low count due to sparsity in dataset therefore those are augmented using SpecAugment \cite{Specaugment}. Since both of the channels are synchronized, corresponding spectrograms are selected from the microphone channel also. The parameters $\theta_1$ and $\theta_2$ of the b-DCCA networks are updated using the following objective
\begin{equation}
\label{b-dcca}
L_{cca} = \text{corr}(f_{\theta_1}(\tilde{X}^{m}), f_{\theta_2}(\tilde{X}^{a}))
\end{equation}
The proposed b-DCCA model is explained in 
Algorithm \ref{algo}.
\begin{figure}[h]
    \centering
    \includegraphics[width=\Imgwidth, angle=90]{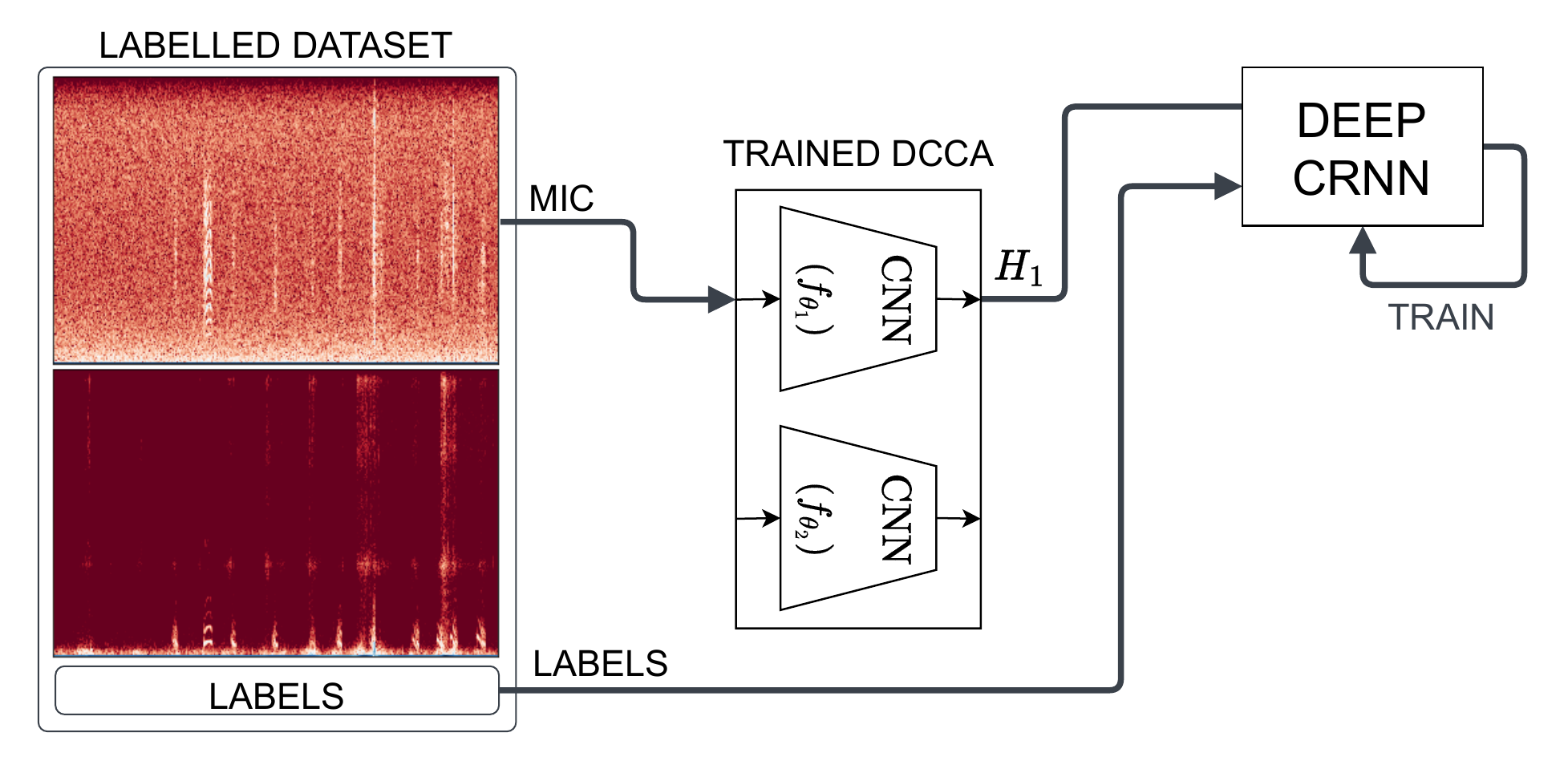}
    \caption{Bird Vocalization Detection Module}
    \label{fig:module3}
\end{figure}
\subsection{Detection}
As shown in the fig.~\ref{fig:module3}, 
the embeddings $H_1$ obtained from trained b-DCCA model is used for bird vocalization detection task using Deep CRNN model as explained in Section \ref{SLM}. 

\begin{algorithm}
 \caption{Algorithm for b-DCCA}
 \begin{algorithmic}[1]
 \label{algo}
 \renewcommand{\algorithmicrequire}{\textbf{Input:}}
 \renewcommand{\algorithmicensure}{\textbf{Output:}}
 \REQUIRE Spectrograms: $(\tilde{X}^m, \tilde{X}^a)\in \{\mathcal{D}_l \cup \mathcal{D}_u\} $ and DCRNN model $f_{\theta}$ trained on $\tilde{X}^a \in \mathcal{D}_l$ 
 \ENSURE Correlated and balanced embeddings: $H_1$
 \\
  \FOR { each $ \tilde{X}^a \in \mathcal{D}_u$ }
   \STATE Detect $\hat{y}^ {a} $ using trained DCRNN model $f_{\theta}$
  \STATE   Calculate $m$ on binarized values of $\hat{y}^a$ such that $m = \sum_t \hat{y}^ {a}$; $t$ = number of time frames in $\tilde{X}^a$
  \STATE Assign a bin number $n$ for each $\tilde{X}^a \in \mathcal{D}_u ~\text{based on} ~m ; ~n = \mathcal{B}(\tilde{X}^{a})$; where $\mathcal{B}$ is binning function as defined in eq.~\ref{bin}
  \STATE Create training batches by sampling uniformly from each bin; $ \tilde{X}_{u}\sim \mathcal{U} (\{\tilde{X}_{u}^{(n)})\} $;~where $n \in \{1, \cdots, B\}$ for $b$ number of bins 
  \STATE Since both the channels are synchronized, sample the corresponding spectrograms $\tilde{X}^m \in \mathcal{D}_u$ 
  \STATE Train b-DCCA $(f_{\theta_1}, f_{\theta_2})$ with the batches created by samples obtained in Step 5, 6
  \ENDFOR
  \STATE Using trained b-DCCA model, obtain the embeddings of  $\tilde{X}^m \in \mathcal{D}_l$  as $H_1 = f_{\theta_1}(\tilde{X}^m)$
 \RETURN $H_1$ 
 \end{algorithmic} 
 \end{algorithm}

 \section{Experiments}
 \label{erd}
 We clip each file into 4-second segments. For calculating the STFT of the segments, we use 1024 point FFT using a 43 ms Hanning window and a hop size of 21.5 ms. The obtained spectrograms have dimension ${(f \times t)}$, where $f=257$ and $t=375$. In all of the three methods discussed below, the performance of the trained classifier (DCRNN) is evaluated on the test data created by random train-test split by sklearn \cite{skl}
 The experiments shown in Table \ref{results} are disussed below.
 

\textbf{DCRNN:} Using the DCRNN \cite{cls} classifier, we classify each time frame of the spectrogram $\tilde{X}^m \in \mathcal{D}_l$ and obtain a vector $\hat{y} \in \mathbb{R}^{t}$. We use 3 Conv2D layers followed by BatchNormalization, relu activation, and Maxpooling2D layers. The features extracted from convolutional layers act as the input to bidirectional GRU. We use single bi-GRU layer followed by dense layer with sigmoid activation. 

$\textbf{DCRNN}^*\textbf{:}$ We use the same architecture and training hyperparameters for training $\text{DCRNN}^*$ model as described above in DCRNN. For $\text{DCRNN}^*$ model, we use accelerometer channel data $\tilde{X}^a \in \mathcal{D}_l$ as input along with the labels. We are not using $\text{DCRNN}^*$ as a baseline because it is trained on accelerometer data therefore performs better; while other models in Table \ref{results} are trained on microphone data.

\textbf{DCCA:} \label{arch} We implement DCCA \cite{b8} baseline using 4 Conv1D layers followed by BatchNormalization and relu activation layer for each view. The DCCA models $f_{\theta_1} ~\text{and}~ f_{\theta_2}$ are trained using $\tilde{X}^m ~\text{and}~ \tilde{X}^a \in \mathcal{D}_u $ respectively. Using trained DCCA, we obtain $(H_1, H_2)$, each having size  $(50 \times 375)$. We select top 50 dimensions along frequency axis while preserving the temporal dimension. For bird vocalization detection task, another DCRNN classifier is trained on labeled data with $H_1$ as input. 

\textbf{b-DCCA:} The architecture and training hyperparameters of the proposed b-DCCA model are similar to that of the DCCA model described above. We train the b-DCCA model using the training batches created by the binning technique as discussed in Section \ref{mod2}. Using trained b-DCCA model for inference, we use $\tilde{X}^m \in \mathcal{D}_l$ to obtain $H_1$. $H_1$ is further used for birds vocalization detection task using the above discussed DCRNN classifier. 



\section{Results}
To evaluate the b-DCCA against the baseline, segment-based F1 score is used as the evaluation metric. 
We use sed\_eval library \cite{eval} for evaluation.

We report P, R, and F1 in Table \ref{results}. The labeled microphone channel of the dataset is used for the sound event detection task. We select microphone channel for bird sound vocalization task because collecting the data from microphone sensor is easy as compared to accelerometer sensor.
The microphone channel contains noisy audio recordings. So, the spectrograms obtained from the microphone channel data does not have high fidelity. Therefore, the supervised DCRNN classifier does not perform very well on this data as reported in Table \ref{results}. The other modality in dataset, i.e., the accelerometer channel is more reliable as compared to microphone channel data. We quantify the fidelity of data by comparing the DCRNN classifier's performance on both of the channels individually. The performance of DCRNN using accelerometer channel data is given in Table \ref{results} as $\text{DCRNN}^*$.

Since, both the microphone and accelerometer channels record the same sound event therefore both of these channels have some hidden correlation. To capitalize on the hidden correlation between both channels, we use DCCA. DCCA extracts useful features from both channels by maximizing the correlation between them. The major issue with DCCA is that due to the high sparsity in data, it learns the solution which is biased towards the majority class, i.e., silence which degrades the results of DCRNN classifier. To address this problem, we applied the proposed b-DCCA on our dataset and demonstrated that the embeddings obtained by the b-DCCA model perform much better than other baseline models.

\begin{figure}[h]
\centering
\includegraphics[scale=0.4]{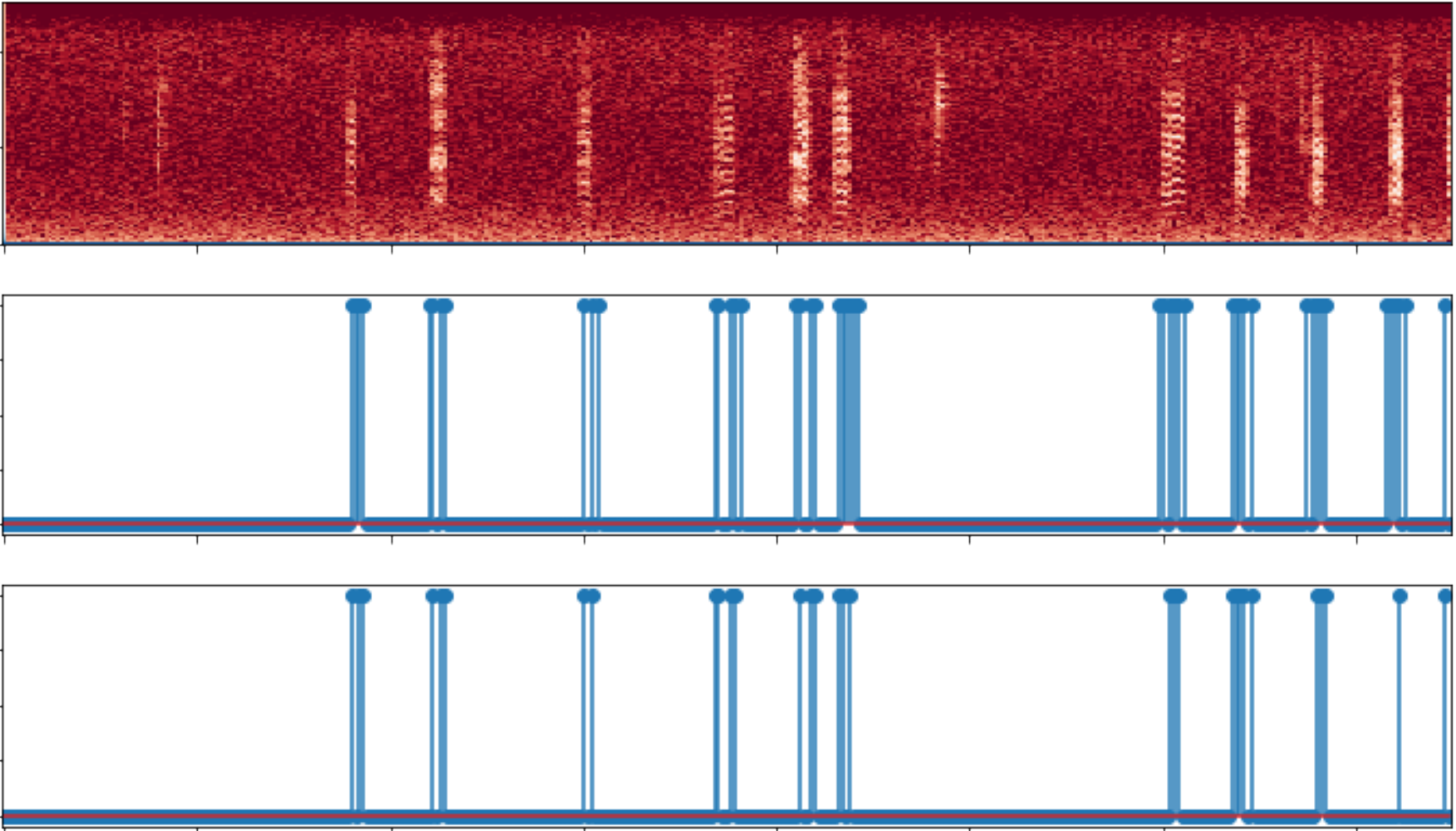}
\caption{Visualization of detected bird vocalizations using b-DCCA model. \textbf{Top:}~Spectrogram of microphone channel data. 
\textbf{Middle:}~ Ground Truth 
\textbf{Bottom:} Predictions}
\label{fig3}
\end{figure}

\begin{table}
\label{tab}
\centering
\caption{Performance of the Proposed b-DCCA Model}
\label{results}
    \begin{tabular}{l c c c}
    
    \hline \hline
    \textbf{Methods} &\textbf{Precision}&\textbf{Recall}&\textbf{F-1 Score}\\\hline
    
    DCRNN \cite{cls} & 0.76 & 0.77 & 0.76  \\
    DCRNN \cite{cls}$^*$ & 0.89 & 0.94 & 0.92 \\
    DCCA \cite{b8} & 0.53 & 0.67 & 0.59  \\
    \textbf{b-DCCA}  &\textbf{0.98} & \textbf{0.72} & \textbf{0.83} \\
    \hline 
    \end{tabular}
\end{table}

\section{Conclusion}
This paper proposes a novel self-supervised algorithm called b-DCCA which uses canonical correlation to learn the hidden relationships between the microphone and accelerometer channel recording the same audio event simultaneously. The obtained results demonstrate that the model generates better embeddings when it learns with a substantial amount of data. Naturally, there is a lot of potential for development. In future works, we hope to move forward with the development of an end-to-end model which performs both of the tasks, i.e., maximizing the canonical correlation between the two views as well as the downstream tasks. Moreover, instead of using sigmoid activation on the final layer of DCRNN classifier, we plan to try other convex functions to make the spectrograms more uniformly distributed across all the bins.

\label{sec:refs}

\bibliography{bibliography}

\end{document}